\documentclass[preprint,12pt]{elsarticle}

\usepackage{amssymb}
\usepackage{amsmath}
\usepackage{bm}
\usepackage[colorlinks,linkcolor=black,anchorcolor=blue, urlcolor=blue,citecolor=blue,unicode]{hyperref}
\usepackage{listings}
\usepackage{subfigure}
\usepackage{graphicx}
\usepackage{tabularx}
\usepackage{array}
\usepackage{booktabs} 
\usepackage{ulem}
\usepackage[version=3]{mhchem}
\usepackage[linesnumbered,ruled,vlined]{algorithm2e}
\usepackage{xcolor}

\newcommand{\bfr}{ {\mathbf r}} 
\newcommand{\bfrp}{ {\mathbf r'}} 
\newcommand{\bfR}{ {\mathbf R}} 
\newcommand{\bfo}{ {\bf 0}} 
\newcommand{\bfq}{ {\mathbf q}} 
\newcommand{\bfk}{ {\mathbf k}} 

\newcommand{\ii}{\mathrm{i}}
\newcommand{\ee}{\mathrm{e}}

\journal{Computer Physics Communications}

\begin{document}

\begin{frontmatter}



\title{LibRPA: A Software Package for Low-scaling First-principles Calculations of Random Phase Approximation Electron Correlation Energy
Based on Numerical Atomic Orbitals}

\author[inst1,inst2]{Rong Shi}

\author[inst2,inst4]{Min-Ye Zhang}
\ead    {minyez@iphy.ac.cn}

\author[inst2,inst3]{Peize Lin}

\author[inst1]{Lixin He}
\ead    {helx@ustc.edu.cn}

\author[inst2]{Xinguo Ren}
\ead{renxg@iphy.ac.cn}

\affiliation[inst1]{
            organization={CAS Key Laboratory of Quantum Information, University of Science and Technology of
            China}, 
            city={Hefei},
            postcode={230026},
            state={Anhui}, 
            country={China}}
            
\affiliation[inst2]{
            organization={Institute of Physics, Chinese Academy of Sciences}, 
            city={Beijing},
            postcode={100190}, 
            country={China}}

\affiliation[inst4]{
            organization={The NOMAD Laboratory at the FHI Molecular Physics Department of the Max-Planck-Gesellschaft}, 
            city={Berlin-Dahlem}, 
            postcode={14195}, 
            state={}, 
            country={Germany}}

\affiliation[inst3]{
            organization={Songshan Lake Materials Laboratory}, 
            city={Dongguan}, 
            postcode={523808}, 
            state={Guangdong}, 
            country={China}}





\begin{abstract}
LibRPA is a software package designed for efficient calculations of random phase approximation (RPA) electron correlation energies from first principles using numerical atomic orbital (NAOs). Leveraging a localized resolution of identity (LRI) technique, LibRPA achieves \(O(N^2)\) or better scaling behavior, making it suitable for large-scale calculation of periodic systems. Implemented in C++ and Python with MPI/OpenMP parallelism, LibRPA integrates seamlessly with NAO-based density functional theory (DFT) packages through flexible file-based and API-based interfaces. In this work, we present the theoretical framework, algorithm, software architecture, and installation and usage guide of LibRPA. Performance benchmarks, including the parallel efficiency with respect to the computational resources and the adsorption energy calculations for \ce{H2O} molecules on graphene, demonstrate its nearly ideal scalability and numerical reliability. LibRPA offers a useful tool for RPA-based calculations for large-scale extended systems.

\section*{Program summary}
\begin{itemize}
\item Program title: LibRPA
\item Developer's repository link: https://github.com/Srlive1201/LibRPA
\item Licensing provisions: LGPL
\item Programming language: C++, Python
\item Nature of problem: Calculating RPA electron correlation energies is computationally expensive, typically scaling as \(O(N^4)\) with system size, hindering its application to large-scale materials science problems.
\item Solution method: LibRPA utilizes the Localized Resolution of Identity (LRI) technique, reducing computational scaling to \(O(N^2)\) or better. Implemented in C++ and Python with MPI/OpenMP parallelization, it integrates with NAO-based DFT packages, facilitating efficient and accurate RPA calculations for large-scale periodic systems.
\end{itemize}

\end{abstract}



\begin{keyword}
First-principles calculations
\sep Random-phase approximation
\sep Numerical atomic orbitals
\sep Resolution of Identity 
\sep Low-scaling algorithm

\end{keyword}

\end{frontmatter}


\section{Introduction}
\label{sec:introduction}

Random phase approximation (RPA) \cite{Bohm/Pines:1953,Gell-Mann/Brueckner:1957,Hubbard:1957b} is a powerful method for calculating the ground-state energy of interacting many-electron systems from first principles \cite{Furche:2001,Fuchs/Gonze:2002,Eshuis/Bates/Furche:2012,Ren/etal:2012b}. 
Formulated within the framework of adiabatic-connection fluctuation-dissipation theorem (ACFDT) \cite{Langreth/Perdew:1977,Gunnarsson/Lundqvist:1976},
RPA can be viewed as a fifth-rung functional according to the Jacob's ladder \cite{Perdew/Schmidt:2001} that classifies different 
density functional approximations \cite{Hohenberg/Kohn:1964,Kohn/Sham:1965}. Compared to lower-rung functionals, RPA captures seamlessly
nonlocal electron correlation effects and provides a balanced description of different chemical bonding interactions, rendering it particularly effective in describing delicate energy differences. Benchmark calculations in the past have shown the remarkable performance of
RPA in describing properties such as surface adsorption energies \cite{Ren/etal:2009,Schimka/etal:2010}, reaction barrier heights \cite{Paier/etal:2012,Ren/etal:2013}, and the energy hierarchy among different polymorphs of crystals \cite{Lebegue/etal:2010,sun:2019,Sengupta/etal:2018,Cazorla/Gould:2019,Yang/Ren:2022} or among different conformers of clusters
\cite{Chedid/Jocelyn/Eshuis:2021,Tahir/etal:2022}. Furthermore, RPA is also a key component of more sophisticated methodologies such
as renormalized second-order perturbation theory \cite{Ren/etal:2013} 
or double hybrid functionals \cite{Zhang/Xu:2019}. Despite its attractive features, 
the application of RPA in materials science has been limited by its high computational cost \cite{eshuis2012electron,Ren/etal:2009,Ren/etal:2012b,ren2012resolution,rocca2012block,Nguyen/deGironcoli:2009,Harl/Kresse:2008,Lu/Nguyen/Galli:2010}, which formally scales as \(O(N^4)\)\cite{harl2009accurate,olsen2013random,rocca2014random,Eshuis/Yarkony/Furche:2010,del2015enabling,ren2012resolution} with respect to system size \(N\).

To reduce the computational cost of RPA for large systems, considerable recent efforts have been devoting to
developing low-scaling algorithms \cite{Neuhauser/etal:2013,Moussa:2014,Kaltak/Klimes/Kresse:2014,Kallay:2015,Wilhelm/etal:2016,Graf/etal:2018,Luenser/Schurkus/Ochsenfeld:2017,Lu/Thicke:2017,Duchemin/Blase:2019}. These include the space-time method that reduces the scaling to \(O(N^3)\) \cite{Rojas/Godby/Needs:1995,White/Godby/Rieger/Needs:1998,Rieger/etal:1999}and the use of localized basis functions combined with the resolution of identity (RI) approximation \cite{Eshuis/Yarkony/Furche:2010,del2015enabling,ren2012resolution}. Most recently, within the numerical atomic orbital (NAO) basis set framework, paired with the localized resolution of identity (LRI) technique \cite{Merlot/etal:2013,Wirz/etal:2017,Hollman/Schaefer/Valeev:2014,Wang/Lewis/Valeev:2020,Levchenko/etal:2015,Lin/Ren/He:2020,Lin/Ren/He:2021}, we have developed an algorithm of formally $O(N^2)$ scaling that works both
for finite systems and periodic systems with finite $\bfk$-point sampling \cite{shi2024subquadratic}. 
Further exploiting the decay properties of the Green's functions in real space, one can achieve 
asymptotically linear scaling for system sizes of approximately 1000 atoms.\cite{shi2024subquadratic}.

Building on these developments, here we introduce a software package called LibRPA, designed for performing efficient and accurate first-principles RPA correlation energy calculations using localized atomic orbitals (AOs). By leveraging the LRI technique, LibRPA achieves \(O(N^2)\) or better scaling behavior for calculating the key quantity in the RPA formalism, i.e., the non-interacting Kohn-Sham (KS) density response function. The software is written in C++ and Python, and it is massively parallelized by means of the MPI/OpenMP hybrid scheme. LibRPA can be integrated with AO-based density functional theory (DFT) packages (provided that the LRI infrastructure is available) via flexible interfaces, making it a versatile tool for computational materials research.

The paper is organized as follows. Section~\ref{sec:theory} presents the theoretical framework behind the LibRPA implementation, which is followed by an exposition of the software architecture and the implementation details in Sec.~\ref{sec:software}. Then, in Sec.~\ref{sec:installation} we describe how the LibRPA package is installed and used in practical calculations. The performance of LibRPA in parallel scalability is reported in
Sec.~\ref{sec:performance}.  A case study of the adsorption of a \ce{H2O} molecule on graphene using LibRPA is presented in Sec.~\ref{sec:application}.
Finally, we summarize this work in Sec.~\ref{sec:summary}. 

\section{Theoretical framework}
\label{sec:theory}
In this section, we briefly recapitulate the key concept and the main equations behind the LibRPA implementation. A full exposition of these equations can be found in
Ref.~\cite{shi2024subquadratic}.

\subsection*{The Localized Resolution of Identity (LRI)}

The localized resolution of identity (LRI) approximation, also known as pair atomic density fitting (PADF) \cite{Merlot/etal:2013, Wirz/etal:2017} or concentric atomic density fitting (CADF) \cite{Hollman/Schaefer/Valeev:2014, Wang/Lewis/Valeev:2020} in the literature, is the key numerical technique that enables low-scaling algorithms in computational chemistry. It has been exploited effectively in various contexts, including periodic hybrid functional calculations \cite{Levchenko/etal:2015, Lin/Ren/He:2020, Lin/Ren/He:2021, Kokott/etal:2024}, $G_0W_0$ computations \cite{Ren/etal:2021}, as well as in calculating the RPA correlation energies \cite{Spadetto/etal:2023} and forces \cite{Tahir/etal:2022}. Benchmark studies have demonstrated its sufficient accuracy for Hartree-Fock exchange evaluations within the NAO framework for both molecular \cite{Ihrig/etal:2015} and periodic systems \cite{Levchenko/etal:2015, Lin/Ren/He:2020, Lin/Ren/He:2021, Kokott/etal:2024}. While the LRI errors are more pronounced in correlated methods involving unoccupied states, high-quality auxiliary basis sets can mitigate these errors, making LRI applicable to correlated methods like RPA and $GW$ for molecules \cite{Ihrig/etal:2015, Tahir/etal:2022} and extended systems \cite{IgorZhang/etal:2019, Ren/etal:2021}. Recently, a computational scheme by Spadetto \textit{et al.} further refines PADF-RPA for large molecules by selectively projecting out the problematic high-lying states that cannot be well described by the local RI scheme \cite{Spadetto/etal:2023}.

In the context of using NAOs as the basis set, the LRI approach essentially consists in approximating the product of two NAOs by a linear combination of auxiliary basis functions (ABFs). This is mathematically expressed as
\begin{align}
    &  \varphi_{i}(\bfr-\bfR_1-{\bm \tau}_I)\varphi_{k}(\bfr-\bfR_2-{\bm \tau}_K)     \nonumber \\
    \approx  &  \sum_{\mu \in I} C_{i(\bfR_1),k(\bfR_2)}^{\mu(\bfR_1)} P_\mu(\bfr-\bfR_1-{\bm \tau}_I)  + 
      \sum_{\mu \in K} C_{i(\bfR_1),k(\bfR_2)}^{\mu(\bfR_2)}P_\mu(\bfr-\bfR_2-{\bm \tau}_K) \nonumber \\
  = & \sum_{\mu \in I} C_{i(\bfo),k(\bfR_2-\bfR_1)}^{\mu(\bfo)} P_\mu(\bfr-\bfR_1-{\bm \tau}_I) + 
          \sum_{\mu \in K} C_{i(\bfR_1-\bfR_2),k(\bfo)}^{\mu(\bfo)} P_\mu(\bfr-\bfR_2-{\bm \tau}_K) \, ,
       \label{eq:LRI}
\end{align}
where \(\varphi_i(\bfr)\) and \(\varphi_k(\bfr)\) denote the NAOs centered on atoms \(I\) and \(K\), located at positions \(\bm{\tau}_I\) and \(\bm{\tau}_K\) within the unit cell, respectively. The ABFs \(P_\mu(\bfr)\) are centered on either atom \(I\) or \(K\), and 
\(C_{i(\mathbf{R}_1),k(\mathbf{R}_2)}^{\mu(\mathbf{R}_1)}\) represent the expansion coefficients of the product of NAOs in terms of the ABFs and 
the lattice vector in parentheses indicates the unit cell to which the basis function belongs. Such a notational system facilitates the usage of
the LRI formalism to periodic systems with finite $\bfk$-point sampling.

This approximation allows for a significant reduction in computational complexity, especially when dealing with the evaluation of two-electron integrals in large systems. By leveraging the locality and compactness of NAOs and the associated sparsity of the expansion coefficients $C$, the LRI technique enables efficient and scalable quantum-mechanical calculations.
\subsection*{Green's Function}

The Green's function is a central quantity in quantum-mechanical many-body physics, describing the propagation behavior of quasiparticles and quasiholes in an interacting system. For RPA calculations, the quantity of interest is the non-interacting Green's function \(G^0\) of the KS system. 
Within the NAO basis-set framework, the real-space imaginary-time expression for the non-interacting Green's function is given by
\begin{equation}
	G^0(\bfr,\bfrp,\ii\tau) = \sum_{i,j}\sum_{\bfR_1,\bfR_2} \varphi_i(\bfr-\bfR_1-{\bm \tau}_I) G^0_{i,j}(\bfR_2-\bfR_1,\ii\tau) \varphi_{j}(\bfrp-\bfR_2-{\bm \tau}_J) \, ,   
	\label{eq:green_func}
\end{equation}
where \(G^0_{i,j}(\mathbf{R},i\tau)\) represents the matrix form of the imaginary-time Green's function expanded in terms of NAO basis, 
\begin{align}
	G^0_{i,j}(\textbf R,\ii\tau)= \left\{
	\begin{array}{ll}
	\displaystyle
	-\frac{\ii}{N_\bfk}\sum_{n,\bfk}f_{n\textbf  k} c_{i,n}(\textbf  k) c_{j,n}^\ast (\textbf  k) \ee^{-\ii\textbf  k \cdot \textbf R} \ee^{-(\epsilon_{n,\textbf  k} - \mu) \tau} 
	 & \tau \le 0 \, , \\
	\displaystyle
	\frac{\ii}{N_\bfk}\sum_{n,\textbf  k}(1-f_{n\textbf  k}) c_{i,n}(\textbf  k) c_{j,n}^\ast (\textbf  k) \ee^{-\ii\textbf  k \cdot \textbf R} \ee^{-(\epsilon_{n,\bfk} - \mu) \tau}  
	& \tau >0 \, .
	\end{array} \right.
	\label{eq:green_mat}
\end{align}
Here, \(c_{i,n}(\mathbf{k})\) are the coefficients of the KS eigenvectors, \(f_{n\mathbf{k}}\) the occupation numbers, \(\epsilon_{n,\mathbf{k}}\) the KS orbital energies, and \(\mu\) the chemical potential. The real-space Green's function will be used to construct the non-interacting
response function matrix, as described below.

\subsection*{Non-Interacting Response Function (\(\chi^0\))}

The response function \(\chi^0\) is an essential quantity for computing the RPA correlation energy and various electronic response properties. 
In real-space and imaginary-time domain, \(\chi^0\) can be expressed in terms of the Green's function as \cite{Rojas/Godby/Needs:1995,White/Godby/Rieger/Needs:1998,Rieger/etal:1999}
\begin{equation}
    \chi^0(\mathbf{r}, \mathbf{r'}, \ii\tau) = -\ii G^0(\mathbf{r}, \mathbf{r'}, i\tau) G^0(\mathbf{r'}, \mathbf{r}, -\ii\tau)\, .
\end{equation}
By employing the expression for the Green‘s function in the basis of NAOs as given in Eq.~\eqref{eq:green_func}, \(\chi^0\) can be expanded in terms of NAOs \, ,
\begin{align}
    \chi^0(\textbf r,\textbf r',\ii\tau)=- \ii \sum_{i,j,k,l}\sum_{\bfR_1,\bfR_2, \bfR_3,\bfR_4} 
    & \varphi_i(\bfr-\bfR_1-{\bm \tau}_I) \varphi_k(\bfr-\bfR_3-{\bm \tau}_K) 
     G^0_{i,j}(\bfR_2-\bfR_1,\ii\tau)  \nonumber \\
    & G^0_{l,k}(\bfR_3-\bfR_4,-\ii\tau) \varphi_{j}(\bfrp-\bfR_2-{\bm \tau}_J)\varphi_{l}(\bfrp-\bfR_4-{\bm \tau}_L) \, .
    \label{eq:chi0_NAOs}
\end{align}

Further applying the LRI approximation leads to a more compact expression of \(\chi^0\) in terms of ABFs, 
\begin{equation}
    \chi^0(\bfr,\bfrp,\ii\tau)=\sum_{\mu \in \mathcal{U},\nu \in \mathcal{V}} \sum_{\textbf{R}_1,\textbf{R}_2}P_\mu(\bfr-\textbf{R}_1-{\bm \tau}_{\mathcal{U}})\chi^0_{\mu,\nu}(\textbf{R}_2-\textbf{R}_1,\ii\tau)P_{\nu}(\bfrp-\textbf{R}_2-{\bm \tau}_\mathcal{V})\, .
    \label{eq:chi0_ABFs}
\end{equation}

By combining Eq.~(\ref{eq:chi0_NAOs}), Eq.~(\ref{eq:chi0_ABFs}) and Eq.~(\ref{eq:LRI}), one finally arrives at the matrix form of \(\chi^0\) in the real-space imaginary-time domain,
\begin{align}
\chi_{\mu,\nu}^0(\textbf{R},\ii\tau) &=-\ii \left[\sum_{i \in \mathcal{U}}\sum_{k,\textbf{R}_1}C_{i(\textbf{0}),k(\textbf{R}_1)}^{\mu (\textbf{0})}\left( \sum_{j \in \mathcal{V}}G_{i,j}(\textbf{R},\ii\tau)  \sum_{l,\textbf{R}_2}C_{j(\textbf{0}),l(\textbf{R}_2-\textbf{R})}^{\nu (\textbf{0})}G_{l,k}(\textbf{R}_1-\textbf{R}_2,-\ii\tau)  \right. \right.\nonumber\\ 
& +\sum_{j \in \mathcal{V}}G^*_{i,j}(\textbf{R},-\ii\tau)  \sum_{l,\textbf{R}_2}C_{j(\textbf{0}),l(\textbf{R}_2-\textbf{R})}^{\nu (\textbf{0})}G^*_{l,k}(\textbf{R}_1-\textbf{R}_2,\ii\tau)\nonumber\\
& +\sum_{j \in \mathcal{V}}G_{j,k}(\textbf{R}_1-\textbf{R},-\ii\tau)\sum_{l,\textbf{R}_2}C_{j(\textbf{0}),l(\textbf{R}_2-\textbf{R})}^{\nu (\textbf{0})}G_{i,l}(\textbf{R}_2,\ii\tau)\nonumber\\
&\left.\left. +\sum_{j \in \mathcal{V}}G^*_{j,k}(\textbf{R}_1-\textbf{R},\ii\tau)\sum_{l,\textbf{R}_2}C_{j(\textbf{0}),l(\textbf{R}_2-\textbf{R})}^{\nu (\textbf{0})}G^*_{i,l}(\textbf{R}_2,-\ii\tau) \right) \right] \, .
\end{align}

The derivation of the above equation takes advantage of the symmetry attributes of the expansion coefficients along with index permutation techniques. For an in-depth explanation of these derivations, please refer to Ref.~\cite{shi2024subquadratic}.

The evaluation of $\chi^0_{\mu,\nu}(\textbf{R}, \ii\tau)$ represents the computational bottleneck in the conventional algorithm. The essential point of our work is to reduce the computational scaling of evaluating $\chi^0_{\mu,\nu}(\textbf{R}, \ii\tau)$, which is crucial for the overall efficiency of LibRPA. To this end, we have designed an algorithm based on atomic pair assignments which is illustrated in Algorithm~\ref{alg:algorithm_N2}. 

\begin{algorithm}[H]
    \footnotesize
    \caption{Loop structure of evaluating $\chi^0_{\mu\nu}(\bfR,\ii\tau)$. $\langle\mathcal U(\textbf 0),\mathcal V(\textbf R)\rangle$ denotes
    an atomic pair with the atom ${\mathcal U}$ in the original unit cell $\bfo$ and the atom ${\mathcal V}$ in the unit cell $\bfR$. The symbol $\mathcal N[\mathcal U]$ represents the set of neighboring atoms around the atom $\mathcal U$, and $K(\textbf R_1) \in \mathcal{N}[\mathcal U(\textbf 0)]$ means
    that the atom $K$ (in the unit cell $\bfR_1$) belongs to the neighborhood of the atom $\mathcal U$ (in the original unit cell).}
    \label{alg:algorithm_N2}
    \ForAll { $<\textbf{R},\tau>$}{
        \ForAll { $<\mathcal{U}(\textbf{0}), \mathcal{V}(\textbf{R})>$}{
            \ForAll { $L(\textbf{R}_2) \in \mathcal{N}[\mathcal{V}(\textbf{R})]$ }{
                Calculate $X^\nu_{i,j}(\textbf{R},\ii\tau) = \displaystyle\sum_{l \in L, \textbf{R}_2} C_{j(\textbf{0}), l(\textbf{R}_2 - \textbf{R})}^{\nu (\textbf{0})} G_{i, l}(\textbf{R}_2, \ii\tau)$\;
            }
            \ForAll { $K(\textbf{R}_1) \in \mathcal{N}[\mathcal{U}(\textbf{0})]$ }{
                \ForAll { $L(\textbf{R}_2) \in \mathcal{N}[\mathcal{V}(\textbf{R})]$ }{
                    Calculate $N^\nu_{j,k}(\textbf{R}_1, \textbf{R}, \ii\tau) = \displaystyle\sum_{l \in L, \textbf{R}_2} C_{j(\textbf{0}), l(\textbf{R}_2 - \textbf{R})}^{\nu (\textbf{0})} G_{l, k}(\textbf{R}_1 - \textbf{R}_2, -\ii\tau)$\;
                }
                Calculate $M^\nu_{i,k}(\textbf{R}_1, \textbf{R}, \ii\tau) = \displaystyle\sum_{j \in \mathcal{V}} G_{i, j}(\textbf{R}, \ii\tau) N^\nu_{j,k}(\textbf{R}_1, \textbf{R}, \ii\tau)$\;
                Calculate $Z^\nu_{i,k}(\textbf{R}_1, \textbf{R}, \ii\tau) = \displaystyle\sum_{j \in \mathcal{V}} G_{j, k}(\textbf{R}_1 - \textbf{R}, -\ii\tau) X^\nu_{i,j}(\textbf{R}, \ii\tau)$\;
                Calculate $O^\nu_{i,k}(\textbf{R}_1, \textbf{R}, \ii\tau) = M^\nu_{i,k}(\textbf{R}_1, \textbf{R}, \ii\tau) + M^{\nu*}_{i,k}(\textbf{R}_1, \textbf{R}, -\ii\tau) + Z^\nu_{i,k}(\textbf{R}_1, \textbf{R}, \ii\tau) + Z^{\nu*}_{i,k}(\textbf{R}_1, \textbf{R}, -\ii\tau)$\;
                Calculate $\chi^0_{\mu, \nu}(\textbf{R}, \ii\tau) \mathrel{+}= C_{i(\textbf{0}), k(\textbf{R}_1)}^{\mu (\textbf{0})} O^\nu_{i,k}(\textbf{R}_1, \textbf{R}, \ii\tau)$\;
            }
        }
    }
\end{algorithm}
The core of our algorithm involves breaking down the evaluation of $\chi^0_{\mu,\nu}(\textbf{R}, \ii\tau)$ into manageable blocks associated with individual atomic pairs. The evaluation process starts by iterating over time grid points ${\tau_j}$ and lattice vectors ${\textbf{R}}$ within the Born-von Karman (BvK) supercell. For each $(\tau, \textbf{R})$ pair, the response function matrix is partitioned into blocks corresponding to individual atomic pairs $\langle \mathcal{U}(\textbf{0}), \mathcal{V}(\textbf{R}) \rangle$.  By computing these blocks separately for each atomic pair and assembling them, we retrieve the full $\chi^0_{\mu,\nu}$ matrix. for a given $\textbf{R}$, the number of such atomic pairs 
is $N_{\rm at}^2$, with $N_{\rm at}$ being the number of atoms in a unit cell.

As illustrated in Algorithm~\ref{alg:algorithm_N2}, within the loop over atomic pairs, we run through the atoms $K$ in the unit cell of $\textbf{R}_1$ and $L$ in the unit cell of $\textbf{R}_2$ to compute the intermediate quantities like $N^\nu_{j,k}(\textbf{R}_1, \textbf{R}, \ii\tau)$ and $X^\nu_{i,j}(\textbf{R}, \ii\tau)$ (defined in Algorithm~\ref{alg:algorithm_N2}, and further details can be found in Ref.~\cite{shi2024subquadratic}). The crucial point is that the atoms $K$ and $L$ must be neighbors of $\mathcal{U}(\textbf{0})$ and $\mathcal{V}(\textbf{R})$, respectively. Otherwise, the LRI expansion coefficients are negligible, so that the contribution from distant atoms become insignificant. The number of neighboring atoms is determined by the cutoff radii of NAO basis functions and does not increase with system size, leading to a computational cost that approaches a constant for each block of the response function matrix associated with an atomic pair. Consequently, the total computational cost scales as $O(N_{\rm at}^2 N_\bfR N_\tau)$, which is effectively $O(N^2)$ for evaluating the response function matrix. If one further takes into account of the spatial decay behavior of
the Green's function, the range of the lattice vector $\bfR$ will also be limited, and then the algorithm becomes asymptotically $O(N)$ \cite{shi2024subquadratic}.




\subsection*{RPA Correlation Energy (\(E_c^\text{RPA}\))}

So far, we have developed the algorithm to compute the real-space imaginary-time response-function matrix represented in terms of ABFs. However, for the calculation of the RPA correlation energy, it is more convenient to transform the response function $\chi^0$ to the $\mathbf{k}$ space and the imaginary-frequency domain. Then, the computation of the RPA correlation energy is executed through an integral across the imaginary frequency spectrum, which incorporates the non-interacting response function \(\chi^0\) and the Coulomb potential \(v\),
\begin{equation}
    E^\text{RPA}_c=\frac{1}{2\pi}\int^{\infty}_0 \mathrm{d}\omega\, \text{Tr}\left[\ln(1-\chi^0(\ii\omega)v)+\chi^0(\ii\omega)v\right]\, ,
\end{equation}
where the trace is taken over the matrix form of 
 \(\chi^0(\ii\omega)\) and \(v\), which are usually represented in terms of the ABF basis set. 

The transformation naturally involves two steps.
In the first step, \(\chi^0\) is converted from the imaginary-time domain to the imaginary-frequency domain, which can be achieved through a cosine Fourier transform due to the symmetry properties of \(\chi^0\) \cite{Kaltak/Merzuk/Kresse:2014-low-scalng,Kaltak/etal:2014b,Rieger/etal:1999},
\begin{equation}
    \chi^0_{\mu,\nu}(\mathbf{R}, \ii\omega_k)=-\ii\sum_{j=1}^{N_\tau} \gamma_{jk}\chi^0_{\mu,\nu}(\mathbf{R} , \ii\tau_j)\, \cos(\tau_j \omega_k)\, .
\end{equation}
Here, \(\gamma_{jk}\) are coefficients determined through an \(L^2\) minimization procedure, and the use of nonuniform grids for \(\tau_j\) and \(\omega_k\) from the GreenX library\cite{CP2Kweb,GreenXweb,azizi2023time} ensures an efficient and accurate representation of the frequency-dependent response. The second step consists in the Fourier transform from real space to reciprocal space, which is straightforward due to the periodicity of the system:
\begin{equation}
    \chi^0_{\mu,\nu}(\mathbf{q}, \ii\omega)=\sum_{\mathbf{R}} \ee^{\ii\mathbf{q} \cdot\mathbf{R}}\chi^0_{\mu,\nu}(\mathbf{R},\ii\omega) \, .
\end{equation}

Now, with the \(\chi^0\) matrix ready in the reciprocal space and imaginary-frequency domain, we can proceed to compute the RPA correlation energy. The Coulomb matrix in the basis of ABFs, crucial for this computation, is given by:
\begin{equation}
    V_{\mu\nu}(\bfq) = \sum_{\bfR} \ee^{\ii\bfq\cdot\bfR} V_{\mu\nu}(\bfR) = \sum_{\bfq} \ee^{\ii\bfq\cdot\bfR} \iint \frac{P_\mu(\bfr-{\bm\tau}_{\cal U})P_\nu(\bfrp-{\bm\tau}_{\cal V}-\bfR)}{|\bfr-\bfrp|} \textrm{d}\bfr \textrm{d}\bfrp \, ,
    \label{eq:Coulomb_matr_kspace}
\end{equation}
which represents the Fourier transform of the real-space Coulomb matrix elements between ABFs. Combining the transformed \(\chi^0\) and the Coulomb matrix, we introduce an intermediate quantity, \(\Pi(\mathbf{q}, \ii\omega)=\chi^0(\mathbf{q}, \ii\omega)V(\mathbf{q})\), to facilitate the computation of \(E_c^{RPA}\),
\begin{align}
    E^\text{RPA}_\text{c}&=\frac{1}{2\pi}\frac{1}{N_\mathbf{q}}\sum_\mathbf{q} \int^{\infty}_0 \mathrm{d}\omega\, \text{Tr}\left[\ln(1-\Pi(\mathbf{q}, \ii\omega))+\Pi(\mathbf{q}, \ii\omega)\right] \nonumber \\
    &=\frac{1}{2\pi} \frac{1}{N_\mathbf{q}}\sum_\mathbf{q} \int^{\infty}_0 \mathrm{d}\omega\, \ln\left[\det(1-\Pi(\mathbf{q}, \ii\omega))\right]+\text{Tr}[\Pi(\mathbf
    {q}, \ii\omega)]\, , 
\end{align}
where the property $\textrm{Tr}[\textrm{ln}(A)]=\textrm{ln}[\textrm{det}(A)]$ is used.
The integration over the imaginary frequency domain, coupled with the summation over the $\bfk$ points in the Brillouin zone, yields the RPA correlation energy per unit cell.

\section{Software Architecture}
\label{sec:software}

\subsection{Architecture}
\begin{figure}[!h]
	\centering
	{
        \includegraphics[width=0.8\textwidth]{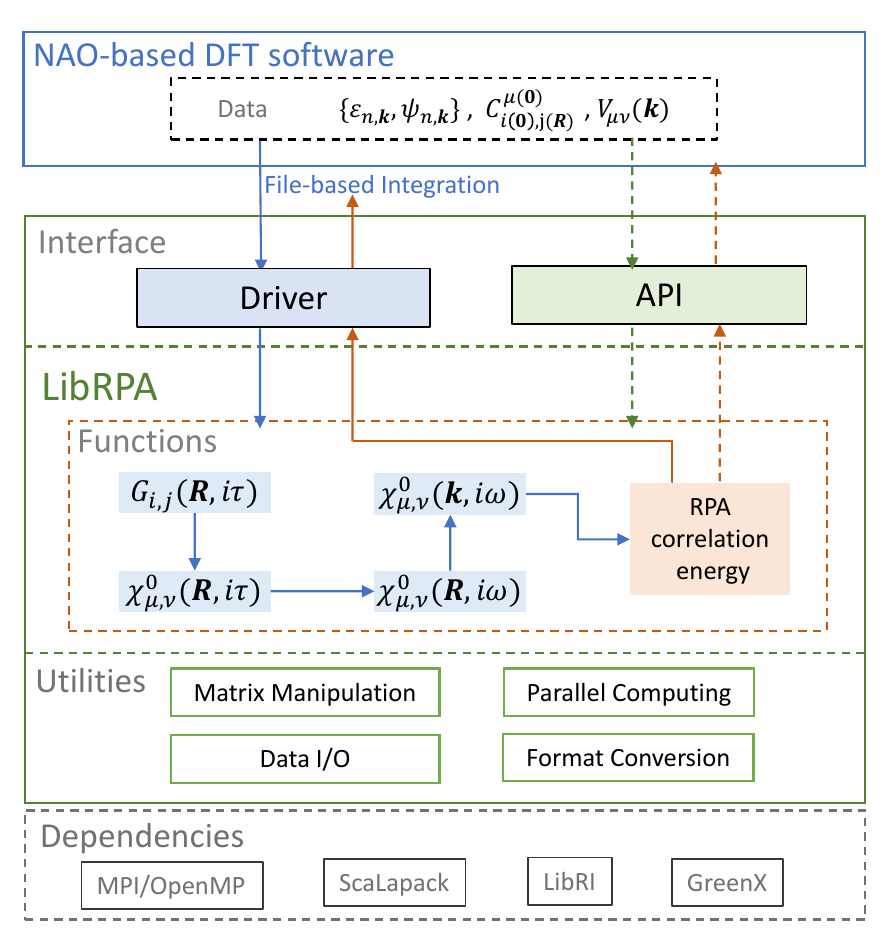} 
    }
      \caption{The architecture of LibRPA. The LibRPA package contains: 1) An interface which allows to integrate the LibRPA with external DFT software via file-based or API methods;
      2) core computational layer that performs the actual RPA calculations, including response function and correlation energy;
      3) the utilities layer that provides matrix manipulation, parallel computing, data I/O, and format conversion tools;
      and 4) dependencies layer that links external libraries like MPI/OpenMP, ScaLapack, LibRI, and GreenX for optimized performance.}   
     \label{fig:LibRPA_stru}
\end{figure}

LibRPA is an independent computer program designed for the implementation of the real-space low-scaling RPA algorithm within the NAO
basis set framework. Fig.~\ref{fig:LibRPA_stru} presents the overall architecture of LibRPA, illustrating its interaction with external DFT software 
and its efficient internal computational process.
At the interface level, LibRPA can be used in two ways, namely, file-based integration through standalone program and  integration through application programming interface (API), depending on how it interacts with external DFT software.
In the former case, LibRPA runs as a post-processing program (driver) that reads data files produced by a DFT software and performs subsequent RPA calculations. In the latter case, LibRPA functions, implemented in a library file (such as \texttt{librpa.so} on Linux system), are called by a DFT software for runtime data transfer and processing.

The core computational layer of LibRPA encapsulates a series of highly optimized algorithmic functions that are pillars of RPA computations. They are responsible for performing various complex mathematical operations, including the evaluation of the response function and the computation of the RPA correlation energy, ensuring both sufficient accuracy and high efficiency.

The utilities layer provides essential matrix manipulation and parallel computing services, supporting LibRPA's high-performance operation. These tools are developed for efficient processing of large datasets and enhancing the management efficiency of multithreading computing tasks, which are crucial for ensuring the high overall computational speed and efficiency of LibRPA.

Finally, the layer of dependencies contains a suite of external libraries required for LibRPA's operation, such as MPI/OpenMP and Scalapack,  providing necessary parallel computing support and optimized numerical operation.  Additionally, the GreenX library 
is employed for generating imaginary frequency/time integration grids \cite{GreenXweb,azizi2023time}, and the LibRI library (optional for RPA correlation energy) can be used for efficient tensor contractions \cite{LibRIweb}. The integration of these libraries ensures that LibRPA can run stably on various hardware and operating system platforms, while efficiently utilizing the computational power of modern multicore processors.

In summary, LibRPA's architecture design balances high efficiency and flexibility, offering a top-tier RPA  computational solution, whether running as an independent software or closely integrated with DFT softwares.

\subsection{Massive parallelization}
LibRPA is massively parallelized, leveraging the power of MPI/OpenMP hybrid parallelism to accelerate the computational tasks.
Throughout the LibRPA workflow, the calculation of the non-interacting response function matrix \(\chi^0\) represents the most time-consuming and critical component. To streamline this process, we have implemented two task distribution schemes, which are chosen depending on the system size and the density of $\bfk$ points.

To explain the parallelization schemes, we define two sets of indices associated with the \(\chi^0\): \({\cal P}\) represents all possible atomic pairs \((\mathcal{U,V})\), where the number of such pairs is given by \(N_{\text{atom}}(N_{\text{atom}}+1)/2\) for a system with \(N_{\text{atom}}\) atoms in the unit cell. \({\cal Q}\) represents all possible combinations of the lattice vectors (\(N_{\bfR}\)) in the BvK supercell and time grid points (\(N_{\tau}\)). The complete response function can be represented as a union of all computed \(\chi^0_{\mathcal{U,V}}(\bfR, \tau)\) values across the atomic pairs \((\mathcal{U,V})\) in \({\cal P}\) and the \((\bfR, \tau)\) grid points in \({\cal Q}\):

\begin{equation}
    \chi^0_{\text{total}} = \bigcup_{(\mathcal{U,V})\in {\cal P}}\;\bigcup_{(\bfR, \tau) \in {\cal Q}} \chi^0_{\mathcal{U,V}}(\bfR, \tau)\;.
\end{equation}

The first scheme, being atom-pair leading, distributes subtasks based on atomic pairs. It partitions the global \(\chi^0\) matrix into blocks associated with these pairs, making it particularly advantageous for large-scale systems. In this context, each parallel subtask is responsible for building the block matrices for all \((\bfR, \tau)\) points associated with a given atomic pair, thus optimizing the distribution of computational loads across multiple processors. Notably, the atom-pair-leading disribution scheme does not require communication until the calculation of the \(\Pi\) matrix is complete.

Alternatively, the second scheme, being \((\bfR, \tau)\)-leading, is optimized for smaller systems or those with a densely sampled Brillouin zone. It allocates tasks based on \((\bfR, \tau)\) points, whereby each parallel subtask calculates the whole \(\chi^0_{\mu,\nu}\) matrix  on specific grid points. In this scheme, communication is required during the Fourier transform. However, since the \(\Pi\) matrix for each \((\bfR, \tau)\) point is complete (not distributed) on a single process, the LU decomposition can be performed independently without further communication.

These two parallelization schemes, different in their data distributions and the requirement of interprocess communications, are designed to address the computational demands in different scientific scenarios. In practice, they are automatically chosen by LibRPA based on specific system characteristics, and by so doing, one can maximize the use of parallel computing resources, ensure efficient load balancing, and minimize interprocess communication overhead. The flow diagrams in \ref{sec:parallel_diagram:appendix} provide a more detailed explanation of the distinct subtask distributions and the specific points at which inter-process communication occurs within the two parallelization schemes.

LibRPA further enhances computational efficiency through OpenMP, which refines the subtasks assigned to MPI processes by dynamically distributing them among threads. This finer subtask granularity utilizes the shared memory capabilities of OpenMP, reducing overall memory consumption while maintaining computational throughput. By optimizing memory usage across threads, OpenMP enables LibRPA to handle large-scale computations more efficiently, ensuring that resources are utilized effectively within the multi-threaded environment.

\subsection{Interface}
LibRPA's interface design is geared towards consistency and compatibility across AO-based DFT software packages written in different programming languages. This applies both to the driver and to the API interfaces. 

For the driver interface, we use standardized data formats for input files of essential physical quantities, including the KS eigenvectors and the expansion coefficients \( C_{i(\mathbf{0}),j(\mathbf{R})}^{\mu(\mathbf{0})} \), as well as the Coulomb matrix \( V_{\mu,\nu} \) in the ABF representation.  Export of essential data in standardized format has been implemented in both FHI-aims \cite{Blum/etal:2009} and ABACUS \cite{Li/Liu/etal:2016}. This standardization facilitates consistent data exchange and integration across different DFT software packages.

For integration through API, the API functions are exposed with C linkage and use one-dimensional array pointers for transferring matrices and tensors, as illustrated in \ref{appendix:API}. This makes the binding to other programming languages easy. Furthermore, multiple API functions for the transfer of Coulomb matrix are provided to address varied data distribution and parallel schemes of the hosting programs.
The target quantity, e.g. RPA correlation energy, can be accessed using dedicated function and further processed within the execution of the host program.

\section{Installation and Usage of LibRPA}
\label{sec:installation}
As mentioned above, LibRPA can work in two modes: either as a standalone program or as a dynamic library (\texttt{librpa.so}) to be integrated into DFT software packages that utilize the NAO basis sets. To facilitate its accessibility, we have streamlined the installation and usage process of LibRPA. In the following, we outline the necessary steps to install LibRPA and explain how to incorporate it into a computational workflow. For detailed usage instructions, including concrete steps to configure and execute RPA calculations, we refer the readers to the documentation page of the LibRPA GitHub repository \cite{LibRPAweb}. 

\subsection{Installation}

To embark on the installation, one should first clone the LibRPA repository from GitHub:

\begin{verbatim}
git clone https://github.com/Srlive1201/LibRPA.git
\end{verbatim}

With the LibRPA package available in a local directory, it is imperative to configure the \texttt{cmake.inc} file to align with one's specific computational environment. This configuration is crucial for setting the compiler flags and specifying the paths to dependencies, thereby ensuring compatibility and achieving optimal performance.

The installation proceeds with the CMake-based build system. To compile LibRPA and generate the \texttt{librpa.so} dynamic library, execute the following commands:

\begin{verbatim}
cmake -B build
cmake --build build
\end{verbatim}

\subsection{Usage}

As mentioned above, LibRPA offers dual operational modes: as an independent executable for direct RPA calculations and as a dynamic library for integration with DFT software such as FHI-aims and ABACUS. This versatility enhances its utility in different computational scenarios.

\subsubsection{Standalone Operation}

In its standalone mode, LibRPA employs a driver program to invoke functionalities from \texttt{librpa.so}, allowing it to perform RPA calculations based on external input data files, generated beforehand by NAO-based DFT program. This mode separates the execution of the DFT program and LibRPA, which is particularly convenient for debugging or developing new features within LibRPA.

\subsubsection{Integration within DFT Software}

When integrating LibRPA as a dynamic library and using LibRPA through API calls, it is necessary to configure the DFT software to recognize and utilize \texttt{librpa.so}. This requires adjusting the DFT software's build process and/or setting environment variables to include the path to the LibRPA library.



\section{Performance}
\label{sec:performance}

\begin{figure}[!h]
	\centering
	{
        \includegraphics[width=0.8\textwidth]{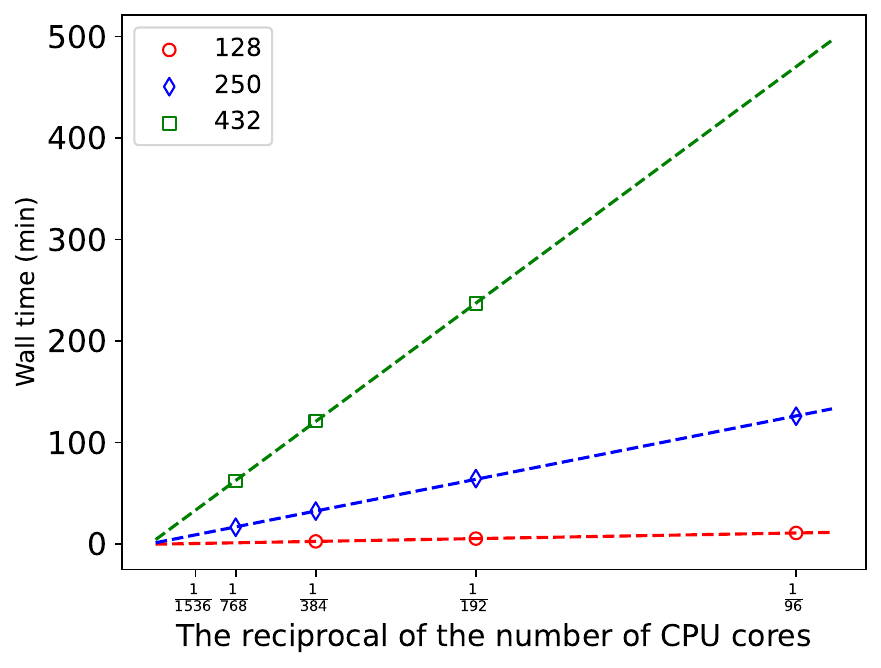} 
    }
      \caption{Strong scaling of the wall time as a function of the number of CPU cores for calculating the RPA correlation energy using LibRPA.
      The test systems are the C diamond supercells of different sizes, containing 128 (red circles), 250 (blue diamonds) and 432 (green squares) atoms. The calculations were performed using Intel Xeon-Silver-4208 CPUs, each with 8 cores, a base frequency of 2.10 GHz, and a maximum frequency of 3.20 GHz.}   
     \label{fig:cpu_atoms_diamond}
\end{figure}

Figure~\ref{fig:cpu_atoms_diamond} presents the strong-scaling behavior of LibRPA with respect to the computational resources. What is plotted are the wall-clock timings for RPA correlation energy calculations for diamond supercells containing 128, 250, and 432 carbon atoms, respectively. The test calculations were performed by interfacing LibRPA with ABACUS using the DZP basis set and a single $\Gamma$-point in the Brillouin-zone sampling. In the LibRPA calculations, 12 minimax time/frequency grid points are used. An ideal scaling of parallelization follows a straight line when plotting the computation time against the reciprocal of the number of CPU cores. Figure~\ref{fig:cpu_atoms_diamond} shows the nearly ideal scaling behavior of LibRPA with respect to computational resources, indicating 
efficient expansion with increased numbers of CPU cores.
For the specific test examples in Fig.~\ref{fig:cpu_atoms_diamond}, the calculation took 237 minutes for the largest system 
(432 atoms), using 192 CPUs (2 nodes, each with 96 CPU cores). Doubling the computing resources to 384 CPUs (4 nodes) reduced the computation time to 121 minutes; and further doubling the resources to 768 CPUs (8 nodes) decreased the time to 62 minutes.  A polynomial fitting of the curve 
of the computation time (in minutes) for the 432-atom test case is \(t = 44736/{N_\text{cores}} + 4.383 (R^2 = 0.9999)\). This 
reflects the excellent balance in task distribution achieved in the design and implementation of the parallel algorithms.

\section{Case Study: \ce{H2O} Adsorption Energy on Graphene}
\label{sec:application}

As a simple case study, we use LibRPA to compute the adsorption energy of a water molecule on graphene. Understanding the adsorption behavior of \ce{H2O} molecules on graphene surfaces is an important theme in studying graphene's surface properties, including the mechanisms behind its hydrophilicity and hydrophobicity \cite{wehling2008first}. Additionally, the adsorption and diffusion behavior of \ce{H2O} molecules on graphene surfaces governs the stability and reactivity of graphene in various environments \cite{hamada2012adsorption, sacchi2023water}. From an application perspective, graphene's unique physicochemical properties make it highly promising in fields such as energy storage, catalysis, and separation technologies. For instance, the adsorption characteristics of \ce{H2O} molecules on graphene directly impact its electrochemical performance when used as an electrode material in supercapacitors and batteries \cite{yang2021electrochemical, sahoo2024recent}.


 A \ce{H2O} molecule can adsorb on graphene surfaces in different configurations, among which the one-leg and two-leg configurations are of particular interest. 
In the one-leg configuration, one \ce{O-H} bond of the water molecule points perpendicularly to the graphene plane, while in the two-leg configuration, the C2 axis of rotation of the water molecule is perpendicular to the plane, with both hydrogen atoms oriented towards the graphene surface.
Lower-rung DFT studies often suggested that the one-leg configuration of water adsorption on graphene is more stable \cite{Ma/etal:2011, hamada2012adsorption}. In contrast, previous RPA calculations have shown that the water molecule binds more strongly with the graphene surface under the two-leg configuration \cite{Ma/etal:2011, brandenburg_physisorption_2019}, which has been verified by experiment \cite{brandenburg_physisorption_2019}. These findings underscore the critical role of RPA in providing accurate and reliable results for water-graphene interactions.  
 


Here we use LibRPA, interfaced with FHI-aims, to calculate the RPA adsorption energies of \ce{H2O} on graphene in the one-leg and two-leg configurations. 
The adsorption energy is defined as
\begin{align}
    \label{eq:adsorption_energy}
    E_{\text{ads}} = E_{\ce{H2O}/\text{graphene}} - E_{\ce{H2O}} - E_{\text{graphene}}\,,
\end{align}
where $E_{\ce{H2O}/\text{graphene}}$, $E_{\text{graphene}}$, and $E_{\ce{H2O}}$ are the total energies of the \ce{H2O}-graphene system, clean graphene, and isolated \ce{H2O} molecule, respectively. In our calculations, a $4\times 4$ graphene supercell was used, with a $4\times 4\times 1$ $\mathbf{k}$-point grid. FHI-aims provides the PBE reference states and corresponding LRI expansion coefficients and Coulomb matrix, while 
the RPA correlation energy is calculated by LibRPA. 
For imaginary frequency and time grids, 16 minimax grid points are used, and for Green's function-based screening, a threshold of $10^{-3}$ is used. These settings provide adequate accuracy.

\begin{figure}[!h]
    \centering
    \subfigure[Two leg ]{
       \includegraphics[width=0.45\textwidth]{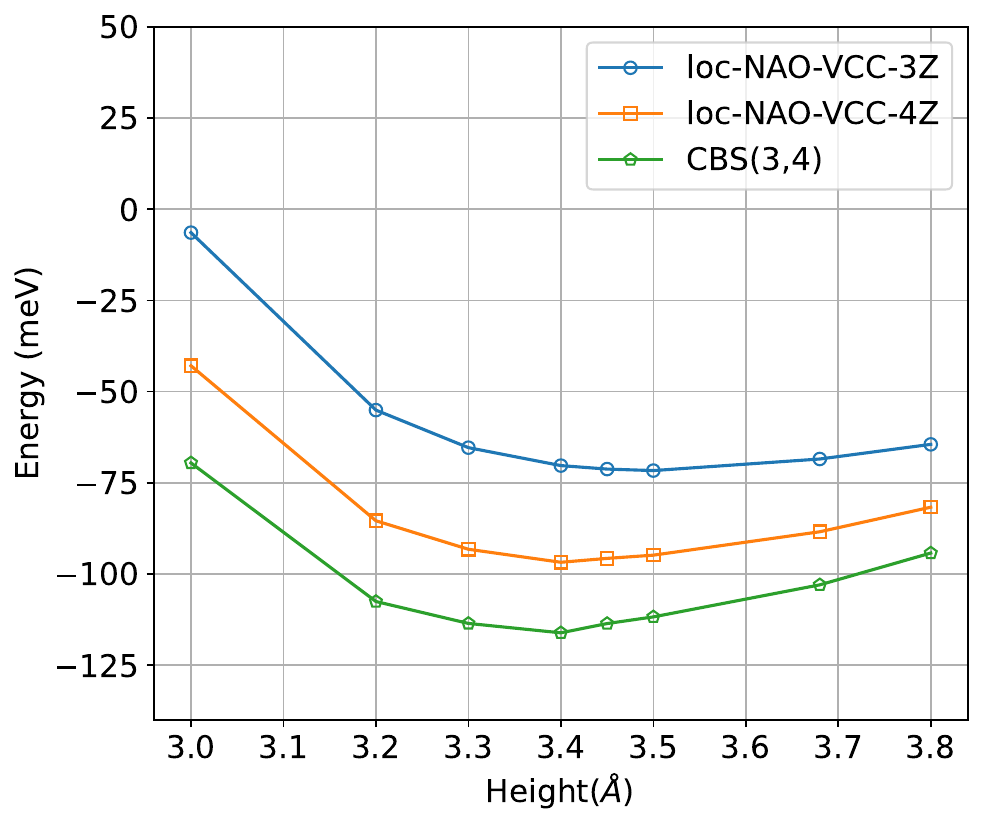}
       \label{fig:graphene_H2O_two}
    }
    \subfigure[One leg ]{
       \includegraphics[width=0.45\textwidth]{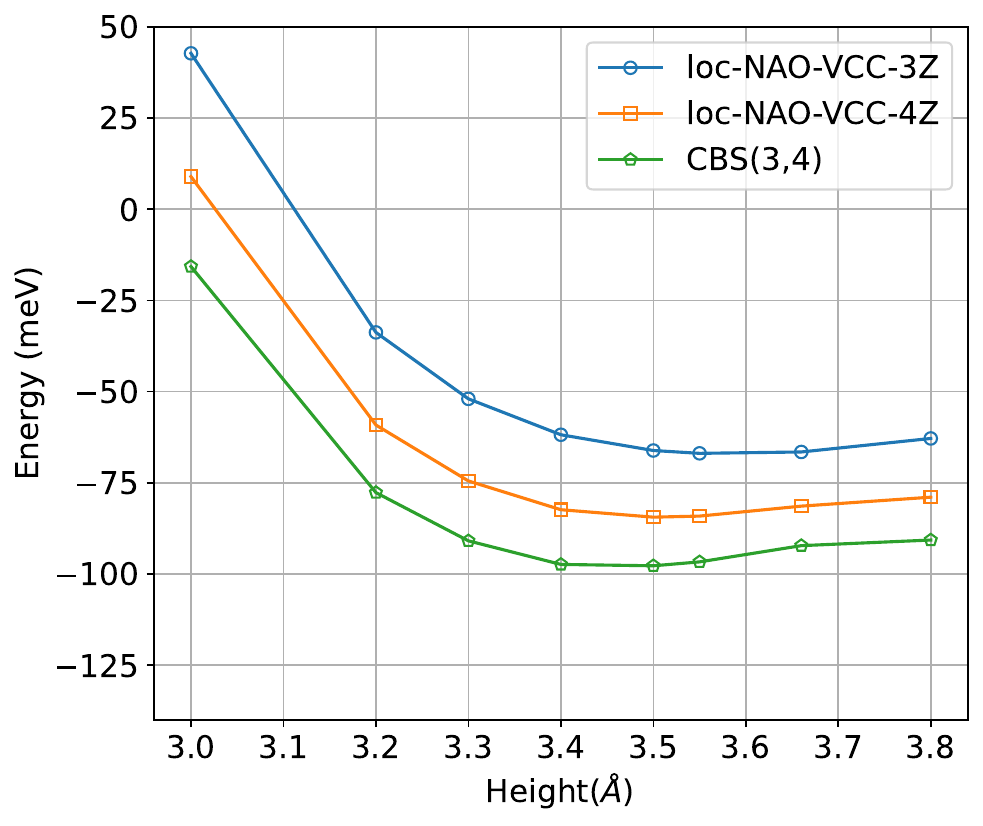}
       \label{fig:graphene_H2O_one}
    }
    \caption{Adsorption energies of a \ce{H2O} molecule adsorbed on graphene as a function of the height of the O atom above the graphene, obtained using LibRPA interfaced with FHI-aims. The left and right panels show the results for the adsorbed water molecule in its two-leg and one-leg configurations, respectively. The three curves correspond to the results obtained using the loc-NAO-VCC-3Z and loc-NAO-VCC-4Z basis sets, and the extrapolated results at the CBS limit using the CBS(3,4) method.}
    \label{fig:graphene_H2O_rpa}
\end{figure}

Figure~\ref{fig:graphene_H2O_rpa} presents the adsorption energy of \ce{H2O} molecule on graphene as a function of the distance between the O atom and the graphene surface under the two-leg (left) and one-leg (right) configurations. The results obtained using the loc-NAO-VCC-3Z and loc-NAO-VCC-4Z  \cite{IgorZhang/etal:2019} basis sets, as well as the extrapolated complete basis set (CBS) results from these two basis sets (denoted CBS(3,4)), are shown. The basis convergence behavior shows that the adsorption energy increases in magnitude (indicating stronger adsorption) as the basis size grows. These results highlight the importance of using high-quality basis sets and the extrapolation method to attain reliable RPA energies. 
Furthermore, we note that the basis set superposition errors (BSSEs) are accounted for in our calculations of the RPA adsorption energies. 
The way in which the BSSEs are corrected in the present calculation is explained in \ref{appendix:bsse_adsorpton_energy}.
Comparing the results in the left and right panels, one finds that the water molecule binds stronger to graphene in the two-leg configuration than in the one-leg configuration, consistent with previous studies \cite{Ma/etal:2011, brandenburg_physisorption_2019}.

\begin{table}[h!]
    \renewcommand{\arraystretch}{1.5}
    \centering
    \caption{Adsorption energy $E_{\text{ads}}$ and height of \ce{H2O} in the two-leg and one-leg configurations on graphene calculated by RPA and
    other methods. The RPA calculations are based on the PBE reference. The plane-wave results are taken from from Ref.~\cite{Ma/etal:2011}, whereby
   the PBE and RPA results were obtained using the VASP code, and the diffusion Monte Carlo (DMC) results are obtained using the CASINO code. }
    
    \footnotesize
    \begin{tabularx}{\textwidth}{c c c c c c}
    \hline\hline
    & & \multicolumn{2}{c}{Two leg} & \multicolumn{2}{c}{One leg} \\
    \cmidrule(lr){3-4} \cmidrule(lr){5-6}
    Approach &Basis set& $E_{\text{ads}}$ (meV) & Height (\AA) & $E_{\text{ads}}$ (meV) & Height (\AA) \\
    \hline
    DMC\textsuperscript{a}  &Plane waves   & $-70 \pm 10$ & 3.4-4.0 & $-70 \pm 10$ & 3.4-4.0 \\
    RPA\textsuperscript{a}  &Plane waves   & $-98 $  & 3.42    & $-82$  & 3.55    \\
    PBE\textsuperscript{a}  &Plane waves  & $-27$        & 3.65    & $-31$        & 3.65    \\
    \\
    \multicolumn{3}{l}{This work} \\
    RPA     &loc-NAO-VCC-3Z  & $-72$      & 3.50    & $-67$  & 3.55    \\
            &loc-NAO-VCC-4Z  & $-97$      & 3.40    & $-84$  & 3.50    \\
            &CBS(3,4)    & $-116$      & 3.40    & $-98$  & 3.50    \\
    PBE     &loc-NAO-VCC-4Z  & $-31$     & 3.68    & $-34$        & 3.66    \\
    \hline\hline
    \textsuperscript{a}Reference \cite{Ma/etal:2011}
    \end{tabularx}
    \label{table:graphene_H2O}
\end{table}

Table \ref{table:graphene_H2O} presents the actual adsorption energies and equilibrium heights of \ce{H2O} molecules on graphene in the two configurations and compares them with the results obtained in a previous study using VASP \cite{Ma/etal:2011}. The table shows that our RPA results using
LibRPA and the loc-NAO-VCC-$n$Z basis sets are in overall satisfactory agreement with those reported in the literature \cite{Ma/etal:2011} with respect to adsorption energies and equilibrium heights. Specifically, the results obtained using loc-NAO-VCC-4Z basis set are in quantitative agreement with the previous plane-wave results, while the CBS(3,4) extrapolated results suggest a stronger binding by an amount of 20 meV in magnitude. This discrepancy might be due to the pseudopotentials used in the plane-wave calculations, or to the remaining uncertainties in the basis set extrapolation scheme, or to the particular procedure for 
correcting  the BSSEs employed in the present work (cf.~\ref{appendix:bsse_adsorpton_energy}). Nevertheless, the present case study demonstrates the capability of LibRPA to be applied to surface
adsorption problems.

\section{Summary}
\label{sec:summary}
In this work, we presented the underlying theoretical foundation and algorithmic details behind the LibRPA software package, which aims to calculate the KS response function and the RPA correlation energy within the AO basis set framework.
In addition, we described the architecture of the LibRPA, the design of its interface with the AO-based DFT softwares, 
its MPI/OpenMP hybrid parallelization scheme, as well as the installation and usage guidelines. Such information is highly useful for the potential users and developers of this software.  

By combining the LRI framework and the spatial locality of the AO basis functions, LibRPA enables 
quadratic-scaling computation of the KS response function matrix, the essential quantity behind the RPA calculations.
Further taking into account of the decaying behavior of the Green's function, LibRPA achieves linear scaling for very large systems.
LibRPA is massively parallelized with the hybrid MPI/OpenMP scheme, and test calculations show its nearly ideal scalability with respect to the computational resources.
To demonstrate the practical usability of LibRPA, we applied it to calculate the adsorption energy and equilibrium height of a \ce{H2O} molecule 
adsorbed on graphene. The results obtained validate the accuracy and reliability of LibRPA, showcasing its capability in handling realistic computational materials science problems.


LibRPA is still under active development, and its capabilities are being continously expanded.
For example, we are currently working on the low-scaling \textit{GW} implementation in LibRPA which supports efficient calculation of the 
quasi-particle energies. Details about the algorithm and implementation of the \textit{GW} module in LibRPA will be described in a future publication.
In short, by utilizing the locality of the AO basis functions and the Green's functions, as well as efficient parallelization schemes,
LibRPA is being developed into a useful tool for performing RPA calculations and beyond.

\appendix
\section{Flow diagrams of the parallel schemes}
\label{sec:parallel_diagram:appendix}
\begin{figure}[!h]
    \centering
    \subfigure[atom-pair leading ]{
        \includegraphics[width=0.5\textwidth]{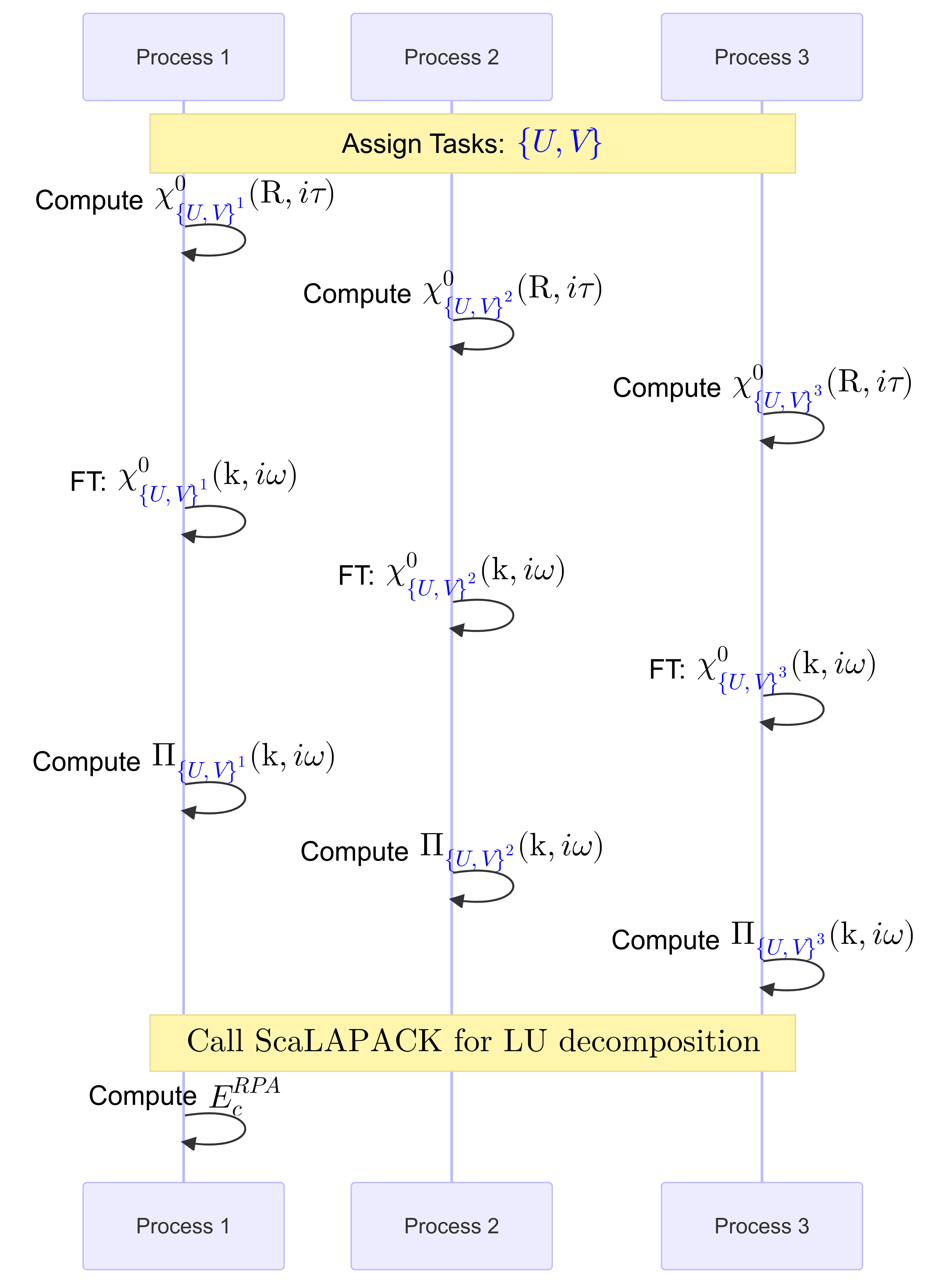} 
        \label{fig:atompair}
    }
    \subfigure[$(\bfR,\tau)$-leading]{
        \includegraphics[width=0.45\textwidth]{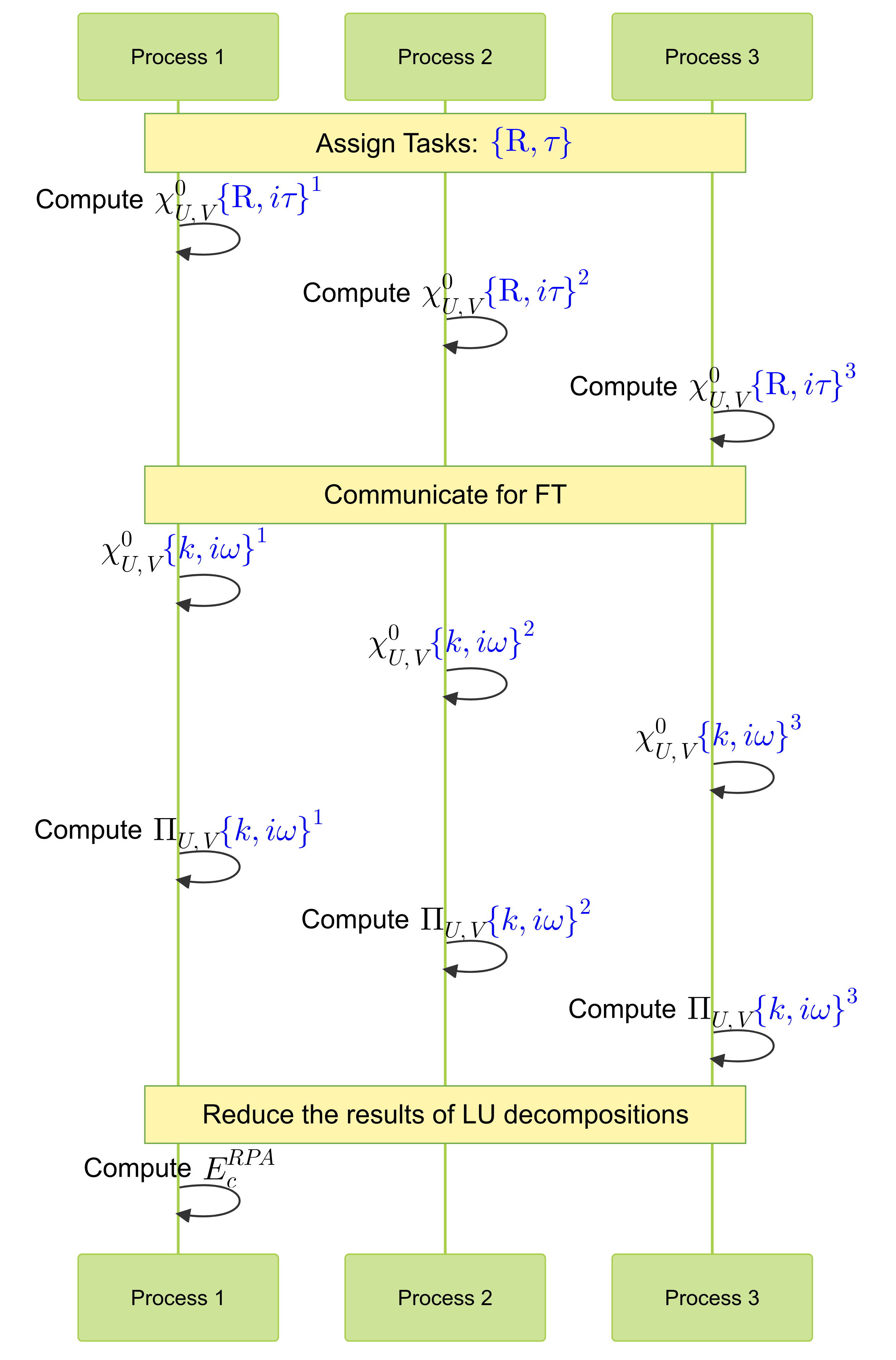} 
        \label{fig:Rtau}
    }
    \caption{Illustration of the parallel distribution schemes implemented in LibRPA: the atom-pair leading parallel scheme (left) and the $(\bfR,\tau)$-leading parallel scheme (right). The distribution of subtasks across processes using different strategies ($\textcolor{blue}{\{\mathcal U, \mathcal V\}}$ or $\textcolor{blue}{\{\bfR, \tau\}}$), which is automatically chosen by LibRPA depending on the type of
    the system to be calculated. The yellow boxes indicate the points at which inter-process communication occurs.}
    \label{fig:parallel_scheme}
\end{figure}

The flow diagrams of the parallel distribution schemes of LibRPA are shown in Fig.~\ref{fig:parallel_scheme}. The left and right panels illustrate
the two distribution schemes based on atomic pairs ${\{\mathcal U, \mathcal V\}}$ and the lattice vector-imaginary time pairs ${\{\bfR, \tau\}}$, respectively. In the first scheme, no communication is needed for the Fourier transform from the real-space imaginary-time domain to
the $\bfk$-space imaginary-frequency domain, but heavy communication is necessary for the LU decomposition of a globally distributed response function matrix. 
On the other hand, in the second scheme, the Fourier transform requires communications across different processes, but the LU decomposition can be
performed locally in each process. 


\section{API functions of LibRPA}
\label{appendix:API}
API functions are provided to allow data exchange between the DFT code and LibRPA at the memory level. This approach eliminates the need for large disk files for input data and enables post-processing of LibRPA output within a single execution of the DFT program. The API functions are categorized into three main groups: environment setup, input parsing and computation. Listing \ref{list:LibRPA_interface} gives a list of API functions as an example. 

  






 





\lstset{language=C,
    breaklines=true,
    basicstyle=\fontsize{11}{12}\sffamily,
    frame=single, 
    rulecolor=\color{black}, 
    framesep=3pt, 
    }
\begin{lstlisting}[caption={API functions of LibRPA}.\label{list:LibRPA_interface}]
void initialize_librpa_environment(
        MPI_Comm comm_global_in, int is_fortran_comm,
        int redirect_stdout, const char *output_filename);

void finalize_librpa_environment();

void set_dimension(int nspins, int nkpts, int nstates, int nbasis, int natoms);

void set_wg_ekb_efermi(int nspins, int nkpts, int nstates, double* wg, double* ekb, double efermi);

void set_ao_basis_wfc(int ispin, int ik, double* wfc_real, double* wfc_imag);

void set_latvec_and_G(double lat_mat[9], double G_mat[9]);

void set_kgrids_kvec_tot(int 
nk1, int nk2, int nk3, double* kvecs);

void set_ibz2bz_index_and_weight(const int nk_irk, const int* ibz2bz_index, const double* wk_irk);

void set_ao_basis_aux(int I, int J, int nbasis_i, int nbasis_j, int naux_mu, int* R, double* Cs_in, int insert_index_only);

void set_aux_bare_coulomb_k_atom_pair(int ik, int I, int J, int naux_mu, int naux_nu, double* Vq_real_in, double* Vq_imag_in);

void set_aux_bare_coulomb_k_2D_block(int ik, int max_naux, int mu_begin, int mu_end, int nu_begin, int nu_end, double* Vq_real_in, double* Vq_imag_in);

void get_rpa_correlation_energy(double *rpa_corr, double *rpa_corr_irk_contrib);
\end{lstlisting}

\section{Correcting the BSSE in adsorption energy calculations}
\label{appendix:bsse_adsorpton_energy}
In RPA calculations of adsorption energies using AO basis sets, a significant challenge is the basis set superposition errors (BSSEs). This error arises from the artificial stabilization due to overlapping basis sets of interacting fragments. While the counterpoise (CP) correction scheme is standard for cluster systems, such a correction scheme is not implemented in FHI-aims for bulk systems. As a workaround, we estimate the magnitude of BSSE based on a cluster model, where the \(\ce{H2O}\) molecule is adsorbed on
the a graphene cluster containing 27 C atoms. We check that further increasing the cluster size does not lead to a noticeable change of
the value of the BSSE. This scheme provides a pragmatic estimate of the BSSE for periodic bulk systems. The results presented in 
Fig.~\ref{fig:graphene_H2O_rpa} and Table~\ref{table:graphene_H2O} are corrected for the BSSE using the above-described approach. 





 \bibliographystyle{elsarticle-num} 
 \bibliography{cas-refs}





\end{document}